\documentclass{ws-procs9x6}

\begin{document}

\title{Three-tangle and Three-$\pi$ for a class of tripartite mixed states}

\author{Teng ma}

\address{Department of Physics, Capital normal University,\\
Beijing, 100048, China\\
E-mail: m.tengteng@163.com}

\author{Shao-Ming Fei}

\address{School of Mathematical Sciences, Capital Normal University,\\
Beijing, 100048, China\\
Max-Planck-Institute for Mathematics in the Sciences,\\
Leipzig, 04103, Germany\\
E-mail: feishm@mail.cnu.edu.cn}

\begin{abstract}

We study the tripartite entanglement for a class of mixed states
defined by the mixture of GHZ and W states,
$\rho=p|GHZ\rangle\langle GHZ|+(1-p)|W\rangle\langle W|$. Based on
the Caratheodory theorem and the periodicity assumption, the
possible optimal decomposition of the states has been derived, which
is not independent on the detailed measure of entanglement. We find
that, according to $p$, there are two different decompositions
containing 3 or 4 quantum states in the decomposition respectively.
When the decomposition contains 3 quantum states, the tripartite
entanglement of the mixed state is simply the entanglement of
superposition states of GHZ and W. When the decomposition contains 4
quantum states, the tripartite entanglement of the mixed state is a
liner function of $p$. We also study the relations between the
three-tangle and three-$\pi$. It is shown that the three-tangle is
smaller than the three-$\pi$. Moreover, the three-$\pi$ has a
minimal point in the interval 0 and 1, while the three-tangle is a
non decreasing function of $p$.

\end{abstract}


\bodymatter

\section{Introduction}\label{aba:sec1}

Quantum entangled states are the key resources in quantum
computation and quantum information processing \cite{Nielsen2000}.
Computation and detection of quantum entanglement are the essential
subjects in the theory of quantum entanglement. For bipartite, in
particular, lower dimensional systems, there are already many useful
results, such as entanglement of formation\cite{Bennett1996,H2001},
concurrece\cite{con}, negativity\cite{G.Vidal},relative
entropy\cite{vv}, PPT criterion\cite{A.P1996} and Bell
inequalities\cite{Bell1996}.

Tripartite entanglement is more complicated than bipartite
entanglement. Investigation of tripartite entanglement is the basis
of studying multipartite entanglement. However, although tripartite
entanglement is well defined, it is formidably difficult to compute
the tripartite entanglement analytically. Up to now, there is no
general formulation to calculate tripartite entanglement, only a few
class of tripartite entanglement can be calculated
efficiently\cite{lohmayer2006, eltschka2007}.

In this article, we study the entanglement of a class of tripartite
mixed stats $\rho  = p\left| {GHZ} \right\rangle \left\langle {GHZ}
\right| + (1 - p)\left| W \right\rangle \left\langle W \right|$. We
give a more detailed impossible optimal decomposition than
\refcite{lohmayer2006}, moreover our formulation can be used not
only three-tangle\cite{coffman}, but also other entanglement
measure. In section 3, we study two important tripartite
entanglement measure and make a comparison between them.

\section{The main formulations}

Consider the tripartite state $\rho=p|GHZ\rangle\langle
GHZ|+(1-p)|W\rangle\langle W|$,
where$|GHZ\rangle=1/\sqrt{2}(|000\rangle+|111\rangle)$,
$|W\rangle=1/\sqrt{3}(|001\rangle+|010\rangle+|100\rangle)$ and
$|0\rangle$, $|1\rangle$ represents the two dimensions of every
partite . If the entanglement of pure states is $E(\Psi_{i})$, then
the entanglement of mixed state is $E(\rho)=min\sum
p_{i}E(\Psi_{i})$\cite{Bennett1996}, where $\rho=\sum p_{i}\Psi_{i}$
and the $min$ is taking all the possible decompositions of $\rho$.
The decomposition
 taking the minimum $\sum p_{i}E(\Psi_{i})$ is call the optimal
 decomposition, and if we know the optimal decomposition of $\rho$,
 then we can get the entanglement $E(\rho)$. One can proof that
 our mixed state $\rho$ can be disassembled by the pure state
 $|q,\theta\rangle=\sqrt{q}|GHZ\rangle-\sqrt{1-q}e^{i\theta}|W\rangle$.

 To get the optimal decomposition of a state, we need to answer two questions:
 first, what is number of pure states of the optimal decomposition?
 second, which is those pure states? To the first question, Caratheodory's theorem\cite{4H} said
 that 4 pure states are sufficient to minimize the entanglement for
 rank-2 states. Hence we need to investigate decompositions with
 2,3, or 4 pure states.

 To answer the second question and consider the symmetry properties of $|GHZ\rangle$ and
 $|W\rangle$ to three parties, we assume that
 the entanglement of $|q,\theta\rangle$ is a periodical function of
 $\theta$ with period $2\pi/3$.
By the assumption that the entanglement of $|q,\theta\rangle$ is a
periodical function of $\theta$ with period $2\pi/3$ for a fixed
$q$, we assume $E(|q,\theta\rangle)$ get minimal value when
$\theta_{n}=\theta_{\ast}+2\pi n/3, n=... -2, -1, 0, 1, 2 ...$. So
$\sum_{i}a_{i}E(q_{i},\theta_{i}\rangle)\geq\sum_{i}a_{i}E(q_{i},\theta_{ni}\rangle)\nonumber$,
then the possible pure state of optimal decomposition becomes
$|q,\theta_{n}\rangle=\sqrt{q}|GHZ\rangle-\sqrt{1-q}e^{i\theta_{n}}|W\rangle$.

We first investigate the optimal decomposition contains 3 pure
states
\begin{equation}
\rho_{op3}=a|q_{1},\theta_{n1}\rangle\langle
q_{1},\theta_{n1}|+b|q_{2},\theta_{n2}\rangle\langle
q_{2},\theta_{n2}|+c|q_{3},\theta_{n3}\rangle\langle
q_{3},\theta_{n3}|
\end{equation}
where $a,b,c\in[0,1]$ and $a+b+c=1$. In fact, the $\rho_{op3}$ also
investigates the situation that the optimal decomposition contains 2
pure stats, because $a,b,c$ can be zero. Under the bases
$\{|GHZ\rangle,|W\rangle\}$, $\rho_{op3}$ and $\rho$ can be
expressed as follows:
 \begin{equation}
 \begin{aligned}
           &\rho_{op3}=\left({\begin{array}{c}
                              aq_{1}+bq_{2}+cq_{3} \\
                              -a\sqrt{q_{1}(1-q_{1})}e^{i\theta_{n1}} -b\sqrt{q_{2}(1-q_{2})}e^{i\theta_{n2}} -c\sqrt{q_{3}(1-q_{3})}e^{i\theta_{n3}}
                            \end{array}}\right.\\
                           &\left.{\begin{array}{c}
                              -a\sqrt{q_{1}(1-q_{1})}e^{-i\theta_{n1}} -b\sqrt{q_{2}(1-q_{2})}e^{-i\theta_{n2}} -c\sqrt{q_{3}(1-q_{3})}e^{-i\theta_{n3}}\\
                              1-(aq_{1}+bq_{2}+cq_{3})
                            \end{array}}
                       \right)\nonumber
\end{aligned}
 \end{equation}
\begin{equation}
\rho_{op3}=\left(
             \begin{array}{cc}
               p & 0 \\
               0 & 1-p \\
             \end{array}
           \right)\nonumber
\end{equation}
Obversely, $\rho_{op3}=\rho$ must hold, we have
\begin{equation}
\left\{ \begin{array}{l}
 aq_1  + bq_2  + cq_3  = p \\
 a\sqrt {q_1 (1 - q_1 )} e^{ - i\theta _{n1} }  + b\sqrt {q_2 (1 - q_2 )} e^{ - i\theta _{n2} }  + c\sqrt {q_3 (1 - q_3 )} e^{ - i\theta _{n3} }  = 0 \\
 a + b + c = 1 \\
 \end{array} \right.
\end{equation}

If $\rho_{op3}$ is the optimal decomposition, then $E(\rho )= aE(|
q_1 ,\theta _{n1}\rangle ) + bE(| q_2 ,\theta _{n2} \rangle ) +
cE(|q_3 ,\theta _{n3}\rangle)$. Because $ aE(|q_1 ,\theta _{n1}
\rangle ),bE(|q_2 ,\theta _{n2}\rangle )$ and $cE(|q_3 ,\theta _{n3}
\rangle) \ge 0$, we have
\begin{equation}\begin{aligned}
aE(|q_1 ,\theta _{n1}\rangle ) + bE(|q_2 ,\theta _{n2}\rangle) +
cE(|q_3 ,\theta _{n3}\rangle) \\ \geq 3\sqrt[3]{aE(|q_1 ,\theta
_{n1}\rangle )bE(|q_2 ,\theta _{n2}\rangle )cE(|q_3 ,\theta
_{n3}\rangle )}\end{aligned}
\end{equation}
 The equality hold if and only if $ aE(|q_1
,\theta _{n1}\rangle ) = bE(|q_2 ,\theta _{n2}\rangle ) = cE(|q_3
,\theta _{n3}\rangle )$, and the definition of $E(\rho)$ demand the
equality of (3) must hold, then we have
\begin{equation}
aE(|q_1, \theta _{n1}\rangle) = bE(|q_2 ,\theta _{n2}\rangle ) =
cE(|q_3 ,\theta _{n3}\rangle)
\end{equation}
Equation (2) and (4) is the conditions that $ a,b,c,q_1 ,q_2 ,q_3$
must satisfy, and we fond (2) and (4) can be satisfied if
\begin{equation}
\left\{{\begin{array}{*{20}c}
   {a = b = c} \\
   {q_1  = q_2  = q_3 }   \\
   {\theta _{n1}  = \theta _* } \\
   {\theta _{n2}  = \theta _*  + \frac{{2\pi }}{3}}  \\
   {\theta _{n3}  = \theta _*  + \frac{{4\pi }}{3}}
\end{array}}\right.
\end{equation}
By (2), (4) and (5), equation (1) becomes
\begin{equation}
\begin{aligned}
&\rho _{opt3}  = \frac{1}{3}\left| {p,\theta _* } \right\rangle
\left\langle {p,\theta _* } \right| + \frac{1}{3}\left| {p,\theta _*
+ \frac{{2\pi }}{3}} \right\rangle \left\langle {p,\theta _*  +
\frac{{2\pi }}{3}} \right| \\&+ \frac{1}{3}\left| {p,\theta _*  +
\frac{{4\pi }}{3}} \right\rangle \left\langle {p,\theta _*  +
\frac{{4\pi }}{3}} \right|
\end{aligned}
\end{equation}
This is the possible optimal decomposition containing 2 and 3 pure
states.

Let us investigate the optimal decomposition that containing 4 pure
states. The optimal decomposition should be (6) adding one pure
state
\begin{equation}
\begin{aligned} &\rho _{opt4}  = a\left| {q',\theta _n ^\prime  } \right\rangle
\left\langle {q',\theta _n ^\prime  } \right| + b(\left| {q,\theta
_* } \right\rangle \left\langle {q,\theta _* } \right| + \left|
{q,\theta _*  + \frac{{2\pi }}{3}} \right\rangle \left\langle
{q,\theta _*  + \frac{{2\pi }}{3}} \right| \\&+ \left| {q,\theta _*
+ \frac{{4\pi }}{3}} \right\rangle \left\langle {q,\theta _*  +
\frac{{4\pi }}{3}} \right|)
\end{aligned}
\end{equation}
where $a,b\in[0,1]$ and $a+3b=1$. Under the bases $ \{ \left| {GHZ}
\right\rangle ,\left| W \right\rangle \}$, $\rho_{opt4}$ can be
expressed as \begin{equation} \begin{aligned}  \rho _{opt4}  =
a\left( {\begin{array}{*{20}c}
   {q'} & { - \sqrt {q'(1 - q')} e^{ - i\theta _n } }  \\
   { - \sqrt {q'(1 - q')} e^{i\theta _n } } & {1 - q'}  \\
\end{array}} \right) + b\left( {\begin{array}{*{20}c}
   {3q} & 0  \\
   0 & {3(1 - q)}  \\
\end{array}} \right)\nonumber
\end{aligned}
\end{equation}
Because $\rho _{opt4}  = \rho$ must hold, we have
\begin{equation}
\left\{ \begin{array}{l}
 aq' + 3bq = p \\
 a\sqrt {q'(1 - q')} e^{ - i\theta _n }  = 0 \\
 a + 3b = 1 \\
 \end{array} \right.
\end{equation}
When $\rho_{op4}$ contains four pure states, $a\neq0$, so $\sqrt
{q'(1 - q')} = 0$, that is $q' = 0$ or $ q' = 1$, so the adding pure
state is $|W\rangle$ or $|GHZ\rangle$. When $q' = 0$, by (8), we get
$b=p/3q$, a=(q-p)/q, due to $a, b\geq0$ and $q>0$, $p$ have an range
$0 \le p \le q$. The corresponding optimal decomposition is $ \rho
_{opt40}  = \frac{{q - p}}{q}\left| W \right\rangle \left\langle W
\right| + \frac{p}{{3q}}(\left| {q,\theta _* } \right\rangle
\left\langle {q,\theta _* } \right| + \left| {q,\theta _*  +
\frac{{2\pi }}{3}} \right\rangle \left\langle {q,\theta _*  +
\frac{{2\pi }}{3}} \right| + \left| {q,\theta _*  + \frac{{4\pi
}}{3}} \right\rangle \left\langle {q,\theta _*  + \frac{{4\pi }}{3}}
\right|)$, and the corresponding entanglement is $E_{opt40}  =
\frac{{q - p}}{q}E\left( {\left| W \right\rangle } \right) +
\frac{p}{q}E\left( {\left| {q,\theta _* } \right\rangle } \right)$.
Note that for a giving $p$, $E_{opt40}$ can vary due to $q$, so $q$
must take a fixed value $q_{*0}$ to make $E_{opt40}$ minimal, then
the final $\rho_{opt40}$ is
\begin{equation}\begin{aligned}
&\rho _{opt40}  = \frac{{q_{*0}  - p}}{{q_{*0} }}\left| W
\right\rangle \left\langle W \right| + \frac{p}{{3q_{*0} }}(\left|
{q_{*0} ,\theta _* } \right\rangle \left\langle {q_{*0} ,\theta _* }
\right| \\&+ \left| {q_{*0} ,\theta _*  + \frac{{2\pi }}{3}}
\right\rangle \left\langle {q_{*0} ,\theta _*  + \frac{{2\pi }}{3}}
\right| + \left| {q_{*0} ,\theta _*  + \frac{{4\pi }}{3}}
\right\rangle \left\langle {q_{*0} ,\theta _*  + \frac{{4\pi }}{3}}
\right|)\end{aligned}
\end{equation}
where $0 \le p \le q$ and $q_{*0}$ is the minimal point of
$E_{opt40}$. When $q' = 1$, by the same method we get $E_{opt41} =
\frac{{p - q}}{{1 - q}}E\left( {\left| {GHZ} \right\rangle } \right)
+ \frac{{1 - p}}{{1 - q}}E\left( {\left| {q,\theta _* }
\right\rangle } \right)$, and
\begin{equation}\begin{aligned}
&\rho _{opt41}  = \frac{{p - q_{*1} }}{{1 - q_{*1} }}\left| {GHZ}
\right\rangle \left\langle {GHZ} \right| + \frac{1}{3}\frac{{1 -
p}}{{1 - q_{*1} }}(\left| {q_{*1} ,\theta _* } \right\rangle
\left\langle {q_{*1} ,\theta _* } \right| \\&+ \left| {q_{*1}
,\theta _*  + \frac{{2\pi }}{3}} \right\rangle \left\langle {q_{*1}
,\theta _*  + \frac{{2\pi }}{3}} \right| + \left| {q_{*1} ,\theta _*
+ \frac{{4\pi }}{3}} \right\rangle \left\langle {q_{*1} ,\theta _* +
\frac{{4\pi }}{3}} \right|)
\end{aligned}\end{equation}
where $q_{*1}  \le p \le 1$ and $q_{*1}$ is the minimal point of
$E_{opt41}$.

Up to now we have investigate all the possible situations,
$\rho_{opt3}$ is the possible decomposition containing two and three
pure states and $\rho_{opt40}$ and $\rho_{opt41}$ is the possible
decomposition containing four pure states. By equations (6),(9),(10)
we can get the corresponding entanglement $E_{opt3}$, $E_{opt40}$,
and $E_{opt41}$, and by the definition of entanglement of
formation\cite{Bennett1996} we have our main formulation
\begin{equation}
E\left( \rho  \right) = \min \{ E_{opt3} ,E_{opt40} ,E_{opt41} \}
\end{equation}
where $E_{opt3}  = E\left( {\left| {p,\theta _* } \right\rangle }
\right)$, when $0 \le p \le 1$, $E_{opt40}  = \frac{{q_{*0}  -
p}}{{q_{*0} }}E\left( {\left| W \right\rangle } \right) +
\frac{p}{{q_{*0} }}E\left( {\left| {q_{*0} ,\theta _* }
\right\rangle } \right)$, when $0 \le p \le q_{*0}$, $E_{opt41} =
\frac{{p - q_{*1} }}{{1 - q_{*1} }}E\left( {\left| {GHZ}
\right\rangle } \right) + \frac{{1 - p}}{{1 - q_{*1} }}E\left(
{\left| {q_{*1} ,\theta _* } \right\rangle } \right)$, when $q_{*1}
\le p \le 1$, and $q_{*0}$, $q_{*1}$ is the minimal point of
$E_{op40}$, $E_{op41}$ respectively.

If we know the entanglement of pure state $|q,\theta\rangle$, and
$E(|q,\theta\rangle)$ is a periodical function of $\theta$ with
period $2\pi/3$, then by equation (11) we can calculate the
entanglement of mixed state $\rho$. If $E_{opt40}$ and $E_{opt41}$
is a differentiable function on $q\in(0,1)$, we get
\begin{equation}
\begin{aligned}
&\frac{{\partial E_{opt40} }}{{\partial q}} = p(\frac{1}{{q^2
}}E(\left| W \right\rangle  - \frac{1}{{q^2 }}E(\left| {q,\theta _*
} \right\rangle ) + q\frac{{dE(\left| {q,\theta _* } \right\rangle
}}{{dq}})\\
&\frac{{\partial E_{opt41} }}{{\partial q}} = (1 - p)( -
\frac{1}{{(1 - q)^2 }}E(\left| {GHZ} \right\rangle ) + \frac{1}{{(1
- q)^2 }}E(\left| {q,\theta _* } \right\rangle )\\&+ \frac{1}{{1 -
q}}\frac{{dE(\left| {q,\theta _* } \right\rangle }}{{dq}})
\end{aligned}
\end{equation}
If $q_{*0}$, $q_{*1}$ is the minimal point of $E_{opt40}$,
$E_{opt41}$ respectively, then $\left. {\frac{{\partial E_{opt40}
}}{{\partial q}}} \right|_{q = q_{*0} }  = 0$, $\left.
{\frac{{\partial E_{opt41} }}{{\partial q}}} \right|_{q = q_{*1} } =
0$, by (12) we known $q_{*0}$, $q_{*1}$ do not depend on $p$. If
$E_{opt40}$ and $E_{opt41}$ have few singular points on $q\in(0,1)$,
we can compare those point with the differentiable points to get the
minimal point.

\section{The study of three-tangle and three-$\pi$ measurement}

Three-tangle\cite{coffman} and Three-$\pi$\cite{yongchengou2007} is
two different important tripartite entanglement measure, in this
section we study these two measurements using our former
formulations. First, let us study the three-tangle. For state
$\left| {q,\theta } \right\rangle = \sqrt q \left| {GHZ}
\right\rangle  - \sqrt {1 - q} e^{i\theta } \left| W \right\rangle$,
the three-tangle is\cite{lohmayer2006}
\begin{equation}
\tau \left( {\left| {q,\theta } \right\rangle } \right) = \left|
{q^2  - \frac{8}{9}e^{i3\theta } \sqrt {6q(1 - q)^3 } } \right|
\end{equation}
Obversely, $\tau(|q,\theta\rangle)$ is a periodical function of
$\theta$ with periodic $2\pi/3$, and when $\theta _n  = \frac{{2\pi
}}{3}n$, $n\in\mathbb{Z}$, the three-tangle of state
$|q,\theta\rangle$ get minimal for a fixed $q$, that is
$\theta_{*}=0$. Note that $\tau_{opt3}$ have a singular point
$q_{*}=\frac{4\sqrt[3]{2}}{3+4\sqrt[3]{2}}\doteq0.627$ on
$q\in(0,1)$, and by (12), we can see that it is also the singular
point of $\tau_{opt40}$ and $\tau_{opt41}$.
 With the equation (11)
we have
\begin{equation}
\tau \left( \rho  \right) = \min \{ \tau _{opt3} ,\tau _{opt40}
,\tau _{opt41} \}
\end{equation}
where $\tau _{opt3}  = \tau \left( {\left| {p,0} \right\rangle }
\right)$, when $0 \le p \le 1$; $\tau _{opt40}  = \frac{p}{{q_{*0}
}}\tau \left( {\left| {q_{*0} ,0} \right\rangle } \right)$, when $0
\le p \le q_{*0}$; $\tau _{opt41}  = \frac{{p - q_{*1} }}{{1 -
q_{*1} }} + \frac{{1 - p}}{{1 - q_{*1} }}\tau \left( {\left| {q_{*1}
,0} \right\rangle } \right)$, when $q_{*1}  \le p \le 1$. To get the
minimal point of $\tau_{opt40}$, we first consider the singular
point $q_{*}$. When $q=q_{*}$, we get $\tau _{opt40}  =
\frac{p}{{q_* }}\tau \left( {\left| {q_* ,0} \right\rangle } \right)
= 0$ and $\tau _{opt41}  = \frac{{p - q_* }}{{1 - q_* }} + \frac{{1
- p}}{{1 - q_* }}\tau \left( {\left| {q_* ,0} \right\rangle }
\right) = \frac{{p - q_* }}{{1 - q_* }}$. $\tau_{opt40}$ has already
get minimal, so $q_{*0}  = q_*$. For $\tau_{opt41}$, we only
consider the interval $q\in(q_{*},1)$ due to when $0 \le p \le 1$,
$\tau_{opt40}$ is already the optimal decomposition. By (12) and
$0<q<1$, we have $2q - 1 > 0$ and $155q^2  - 155q + 32 = 0$, then
get the possible minimal point $q_{*1} ^\prime   = \frac{1}{2} +
\frac{3}{{310}}\sqrt {465} \buildrel\textstyle\over\doteq 0.709$.
Comparing with the singular point $\tau _{opt41} \left( {q_* }
\right) > \tau _{opt41} \left( {q_{*1} ^\prime  } \right)$, so the
minimal point of $\tau_{opt41}$ is $q_{*1}  = q_{*1} ^\prime   =
\frac{1}{2} + \frac{3}{{310}}\sqrt {465}
\buildrel\textstyle\over\doteq 0.709$. By comparing $\tau _{opt3} ,
\tau _{opt40}$ and $\tau _{opt41}$, we get when $0 \le p \le
q_{*0}$, $\tau _{opt40}  \le \tau _{opt3}$, if and only if $p = 0$
or $p = q_{*0}$ equality holds, when $q_{*1}  \le p \le 1$, $\tau
_{opt41}  \le \tau _{opt3}$, if and only if $p = q_{*0}$ or $p = 1$
equality holds, when $q_{*0}  < p < q_{*1}$ there is only
$\tau_{opt3}$. Then by (14) we get
\begin{equation}
\begin{array}{cc}
\tau \left( \rho  \right) = \left\{ \begin{array}{l}
 \frac{p}{{q_{*0} }}( - q_{*0} ^2  + \frac{8}{9}\sqrt {6q_{*0} (1 - q_{*0} )^3 } ) \\
 p^2  - \frac{8}{9}\sqrt {6p(1 - p)^3 }  \\
 \frac{{p - q_{*1} }}{{1 - q_{*1} }} + \frac{{1 - p}}{{1 - q_{*1} }}(q_{*1} ^2  - \frac{8}{9}\sqrt {6q_{*1} (1 - q_{*1} )^3 } ) \\
 \end{array} \right.
&
\begin{array}{l}
 0 \le p \le q_{*0}  \\
 q_{*0}  < q < q_{*1}  \\
 q_{*1}  \le p \le 1 \\
 \end{array}
\end{array}
\end{equation}
where $q_{*0} = {\raise0.7ex\hbox{${4\sqrt[3]{2}}$}
\!\mathord{\left/
 {\vphantom {{4\sqrt[3]{2}} {(3 + 4\sqrt[3]{2})}}}\right.\kern-\nulldelimiterspace}
\!\lower0.7ex\hbox{${(3 + 4\sqrt[3]{2})}$}}
\buildrel\textstyle.\over= 0.627$, $q_{*1}= \frac{1}{2} +
\frac{3}{{310}}\sqrt {465} \buildrel\textstyle.\over= 0.709$, and
(6), (9), (10) is the corresponding optimal decomposition.

To study three-$\pi$, we first need getting the expression of
three-$\pi$ of the pure state $|q,\theta\rangle$. The definition of
three-$\pi$ of a pure state is\cite{yongchengou2007}
\begin{equation}
\pi  = \frac{1}{3}(\pi _a  + \pi _b  + \pi _c )
\end{equation}
where $\pi _a  = N_{a(bc)} ^2  - N_{ab} ^2  - N_{ac} ^2$, $\pi _b =
N_{b(ac)} ^2  - N_{ba} ^2  - N_{bc} ^2$, and $\pi _c  = N_{c(ab)} ^2
- N_{ca} ^2  - N_{cb} ^2$. $N$ is a kind of bipartite entanglement
measure called negativity\cite{4H1998}\cite{G.Vidal}, the definition
is $N_{ab} = \left\| {\rho _{ab} ^{Ta} } \right\| - 1$, $\|\rho\|$
is trace norm, it equals the sum of modulus of eigenvalues of
$\rho$. $\rho _{ab} ^{Ta}$ is the partial transpose of $\rho _{ab}$,
satisfy $(\rho _{ab} ^{Ta} )_{ij,kl}  = (\rho _{ab} )_{kj,il}$. By
pure state $|q,\theta\rangle$, under the natural base, we get
\begin{equation}
\rho _{ab}= \left( {\begin{array}{*{20}c}
   {\frac{q}{2} + \frac{{1 - q}}{3}} & { - \sqrt {\frac{{q(1 - q)}}{6}} e^{ - i\theta } } & { - \sqrt {\frac{{q(1 - q)}}{6}} e^{ - i\theta } } & { - \sqrt {\frac{{q(1 - q)}}{6}} e^{i\theta } }  \\
   { - \sqrt {\frac{{q(1 - q)}}{6}} e^{i\theta } } & {\frac{{1 - q}}{3}} & {\frac{{1 - q}}{3}} & 0  \\
   { - \sqrt {\frac{{q(1 - q)}}{6}} e^{i\theta } } & {\frac{{1 - q}}{3}} & {\frac{{1 - q}}{3}} & 0  \\
   { - \sqrt {\frac{{q(1 - q)}}{6}} e^{ - i\theta } } & 0 & 0 & {\frac{q}{2}}  \\
\end{array}} \right)
\end{equation}
\begin{equation}
\rho _{ab} ^{Ta}  = \left( {\begin{array}{*{20}c}
   {\frac{q}{2} + \frac{{1 - q}}{3}} & { - \sqrt {\frac{{q(1 - q)}}{6}} e^{ - i\theta } } & { - \sqrt {\frac{{q(1 - q)}}{6}} e^{i\theta } } & {\frac{{1 - q}}{3}}  \\
   { - \sqrt {\frac{{q(1 - q)}}{6}} e^{i\theta } } & {\frac{{1 - q}}{3}} & { - \sqrt {\frac{{q(1 - q)}}{6}} e^{ - i\theta } } & 0  \\
   { - \sqrt {\frac{{q(1 - q)}}{6}} e^{ - i\theta } } & { - \sqrt {\frac{{q(1 - q)}}{6}} e^{i\theta } } & {\frac{{1 - q}}{3}} & 0  \\
   {\frac{{1 - q}}{3}} & 0 & 0 & {\frac{q}{2}}  \\
\end{array}} \right)
\end{equation}
and
\begin{equation}
\rho _a= \left( {\begin{array}{*{20}c}
   {\frac{q}{2} + \frac{{2(1 - q)}}{3}} & {\sqrt {\frac{{q(1 - q)}}{6}} e^{ - i\theta } }  \\
   {\sqrt {\frac{{q(1 - q)}}{6}} e^{i\theta } } & {\frac{q}{2} + \frac{{(1 - q)}}{3}}  \\
\end{array}} \right)
\end{equation}
By calculate $\rho_{bc}$, $\rho_{ac}$, $\rho_{b}$, $\rho_{c}$, we
can see that due to the symmetry property of $|GHZ\rangle$ and
$|W\rangle$ to every partite, $\rho _{ab}  = \rho _{bc}  = \rho
_{ac}$, $\rho _a = \rho _b  = \rho _c$, then we have $N_{ab}  =
N_{ac}  = N_{bc}$, $\pi _a = \pi _b  = \pi _c$, $\pi  = \pi _a$. One
can prove for three qubit pure states, $N_{a(bc)}  = C_{a(bc)}$,
where $C_{a(bc)}$ is the $Concurrence$\cite{con} between systems $a$
and $b,c$. Then by (19) we get $N_{ab}=C_{a(bc)}  = \sqrt {2(1 -
Tr\rho _a ^2 )}  = \sqrt {\frac{5}{9}q^2  - \frac{4}{9}q +
\frac{8}{9}}$. By (18) we get the characteristic equation of matrix
$\rho _{ab} ^{Ta}$
\begin{equation}
\begin{array}{l}
 \lambda ^4  - \lambda ^3  + (\frac{5}{{36}}q^2  - \frac{q}{9} + \frac{2}{9})\lambda ^2 \\ + [\frac{(q(1 -
 q))^{3/2}}{3\sqrt 6 }\cos{3\theta}  - \frac{7}{{27}}q^3  + \frac{7}{{18}}q^2  - \frac{q}{6}+ \frac{1}{{27}}]\lambda
  \\+ [ - \frac{q(q(1 -
 q))^{3/2}}{6\sqrt 6 }\cos{3\theta} - \frac{{41}}{{648}}q^4  + \frac{{149}}{{648}}q^3  - \frac{{13}}{{54}}q^2  + \frac{7}{{81}}q - \frac{1}{{81}}] = 0 \\
 \end{array}
\end{equation}
Then the three-$\pi$ of pure state $|q,\theta\rangle$ can be
expressed as
\begin{equation}
\begin{array}{l}
 \pi  = \pi _a  = N_{a(bc)} ^2  - N_{ab} ^2  - N_{ac} ^2  = C_{a(bc)} ^2  - 2N_{ab} ^2  \\
  = \frac{5}{9}q^2  - \frac{4}{9}q + \frac{8}{9} - 2(\sum\limits_{i = 1}^4 {\left| {\lambda _i \left( {q,Cos3\theta } \right)} \right|}  - 1)^2  \\
 \end{array}
 \end{equation}
 \begin{figure}\begin{center}
\psfig{file=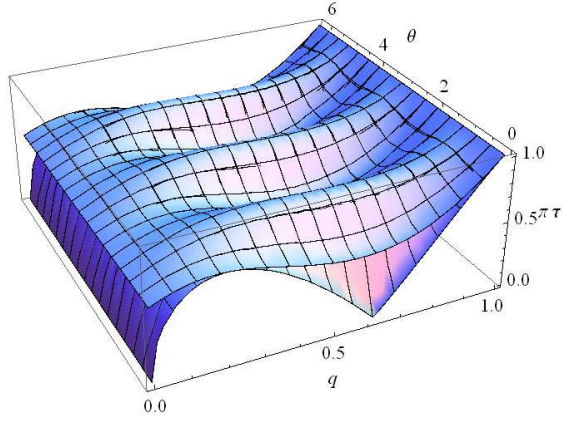,width=3.5in} \caption{The three-$\pi$ and
three-tangle for pure state $|q,\theta\rangle = \sqrt q \left| {GHZ}
\right\rangle  - \sqrt {1 - q} e^{i\theta } \left| W \right\rangle $
as a function of $\theta$ and $q$. The top surface is three-$\pi$,
the bottom surface is three-tangle, and three-$\pi$ is always
greater or equal than three-tangle. For a fixed $q$, three-tangle
and three-$\pi$ are all the periodical functions of $\theta$, and
get minimal when $\theta=0, 2\pi/3, 4\pi/3, ...$}
\label{aba:fig1}\end{center}
\end{figure}
 where $ i = 1,2,3,4$ and $\lambda _i \left( {q,Cos3\theta } \right)$
 is the solutions of (20). From fig1 and (21) we can see that three-$\pi$ is a periodical
 function of $\theta$ with period $2\pi/3$ and when $\theta=2\pi n/3$, $n\in\mathbb{Z}$,
  the three-$\pi$ get minimal, that is $\theta_{*}=0$,
  so we can use our formal
 formulations.
Using our formulation (11) and the method just used for three-tangle
we finial get
\begin{equation}
 \pi \left( \rho  \right) = \left\{ \begin{array}{l}
 \frac{{q_{*0}  - p}}{{q_{*0} }}\frac{4}{9}(\sqrt 5- 1) +\frac{p}{{q_{*0} }}[\frac{5}{9}q_{*0} ^2  - \frac{4}{9}q_{*0}\\ \qquad\qquad +\frac{8}{9}- 2(\sum\limits_{i = 1}^4 {\left| {\lambda _i \left( {q_{*0} ,1} \right)} \right|}  - 1)^2 ] \qquad 0 \le p \le q_{*0}\\
 \frac{5}{9}p^2  - \frac{4}{9}p + \frac{8}{9} - 2(\sum\limits_{i = 1}^4 {\left| {\lambda _i \left( {p,1} \right)} \right|}  - 1)^2 \qquad q_{*0}  \le q \le q_{*1}\\
\frac{{p - q_{*1} }}{{1 - q_{*1} }} +\frac{{1 - p}}{{1 - q_{*1}
}}[\frac{5}{9}q_{*1} ^2  - \frac{4}{9}q_{*1} + \frac{8}{9}\\
\qquad\qquad\qquad - 2(\sum\limits_{i = 1}^4 {\left| {\lambda _i
\left( {q_{*1} ,1} \right)} \right|}  - 1)^2 ] \qquad  q_{*1}  \le p
\le 1
 \end{array} \right.
\end{equation}
where $q_{*0}  = {\rm{0}}{\rm{.564}}...$, $ q_{*1}  =
{\rm{0}}{\rm{.963}}...$, $ i = 1,2,3,4$, $\lambda _i \left({q,1}
\right)$ is the solutions of (20), and (6), (9), (10) is the
corresponding optimal decomposition.

Let us make a comparison between three-tangle and three-$\pi$. We
first consider pure state. For $m\otimes n$, $m\leq n$ bipartite
mixed states, we have $\sqrt {\frac{2}{{m(m - 1)}}} (\left\| {\rho
_{ab} ^{Ta} } \right\| - 1) \le C(\rho _{ab} )$\cite{chenkai}, for
$2\otimes2$ system $N(\rho _{ab} ) \le C(\rho _{ab} )$. For three
qubit pure states we have $N_{a(bc)}  = C_{a(bc)}$, then for three
qubit pure states we have
\begin{equation}
\begin{array}{l}
 \pi  = \frac{1}{3}(\pi _a  + \pi _b  + \pi _c ) = \frac{1}{3}(N_{a(bc)} ^2  - N_{ab} ^2  - N_{ac} ^2  + N_{b(ac)} ^2  - N_{ba} ^2  - N_{bc} ^2 \\ + N_{c(ab)} ^2  - N_{ca} ^2  - N_{cb} ^2 )
  \ge \frac{1}{3}(C_{a(bc)} ^2  - C_{ab} ^2  - C_{ac} ^2  + C_{b(ac)} ^2  - C_{ba} ^2  - C_{bc} ^2 \\ + C_{c(ab)} ^2  - C_{ca} ^2  - C_{cb} ^2 ) = \frac{1}{3}(\tau _a  + \tau _b  + \tau _c ) = \tau
 \end{array}
\end{equation}
Therefore for three qubit pure states, three-$\pi$ is great or equal
than three-tangle. Fig1 shows the entanglement of pure stats
$|q,\theta\rangle$ under these two entanglement measure.

For rank-2 mixed states $\rho  = p\left| {GHZ} \right\rangle
\left\langle {GHZ} \right| + (1 - p)\left| W \right\rangle
\left\langle W \right|$, by (15) and (22) we have fig2. From fig2 we
can see that for this rank-2 class states three-tangle is  smaller
or equal then three-$\pi$, and the two measurements show different
trend when $p$ increases. An interest thing is that on
$p=q_{*0}=0.564...$ three-$\pi$ get minimal $0.50103...$ which
contrary to our intuition that the increase of weight of maximal
entangled state\cite{Gisin} $|GHZ\rangle$ means a lager
entanglement. While, three-tangle is a nondecreasing function of
$p$.
\begin{figure}
\begin{center}
\psfig{file=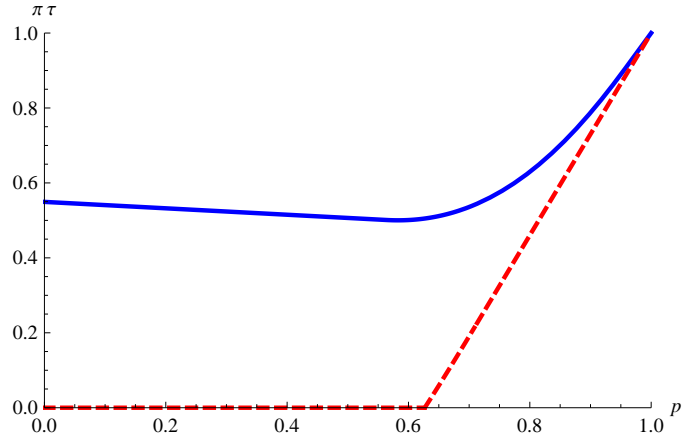,width=3.5in}
\end{center}
\caption{The three-$\pi$ and three-tangle for mixed states $\rho  =
p\left| {GHZ} \right\rangle \left\langle {GHZ} \right| + (1 -
p)\left| W \right\rangle \left\langle W \right|$ as a function of
$p$. The top curve is three-$\pi$, the bottom curve is three-tangle.
We can see that three-tangle and three-$\pi$ are all the liner
functions of $p$ on $(0,q_{*0})$ and $(q_{*1},1)$. For three-tangle,
$q_{*0}=0.627...$, $q_{*1}=0.709...$, and for three-$\pi$,
$q_{*0}=0.564...$, $q_{*1}=0.963...$.} \label{aba:fig1}
\end{figure}

\section{Conclusions}
We study the the entanglement for a class of rank-2 mixed states
$\rho  = p\left| {GHZ} \right\rangle \left\langle {GHZ} \right| + (1
- p)\left| W \right\rangle \left\langle W \right|$. Base on
$Caratheodory$ theorem and the periodicity assumption, the possible
optimal decomposition has been derived. Our optimal decomposition
does not depend on the kinds of entanglement measure if the
entanglement measure satisfy our assumptions. We also apply our
formulation to  study two important tripartite entanglement measure
three-tangle and three-$\pi$. We find three-tangle is always smaller
or equal than three-$\pi$, and three-tangle is a nondecreasing
function of $p$ while three-$\pi$ has a minimal point on
$p\in(0,1)$, and this show that different entanglement measure can
have different trend for the same state. Our study of this class of
tripartite mixed rank-2 states may be useful for studying other
tripartite quantum states and even some multipartite higher
dimensional  states and explore the essence the quantum entanglement
and quantum mechanic.

\section*{Acknowledgments}
The author would like to thank Shao-Ming Fei for the discussions.
\bibliographystyle{ws-procs9x6}

\end{document}